\documentclass[conference]{IEEEtran}
\IEEEoverridecommandlockouts
\usepackage{epstopdf}
\usepackage{array}
\usepackage[caption=false,font=normalsize,labelfont=sf,textfont=sf]{subfig}
\usepackage{textcomp}
\usepackage{stfloats}
\usepackage{url}
\usepackage{verbatim}
\usepackage{graphicx}
\usepackage{cite}
\usepackage[utf8]{inputenc}
\usepackage{booktabs}
\usepackage{amsmath,amssymb,amsfonts}
\usepackage{amsthm}
\usepackage{algorithm}
\usepackage{algpseudocode}
\usepackage{tablefootnote}
\usepackage{graphicx}
\usepackage{tikz}
\usepackage{float}
\usepackage{xcolor}
\usepackage{multirow}
\usepackage{adjustbox}
\usepackage{stfloats}
\usepackage{url}
\usepackage{balance}

\begin{document}

\title{RIS-Assisted Survivable Fronthaul Design in Cell-Free Massive MIMO System}

\author{

 \IEEEauthorblockN{Zhenyu Li$^*$, \"Ozlem Tu\u{g}fe Demir$^\dagger$, Emil Bj{\"o}rnson$^*$, Cicek Cavdar$^*$}
\IEEEauthorblockA{ {$^*$Department of Computer Science, KTH Royal Institute of Technology, Kista, Sweden}
\\ 
 {$^\dagger$Department of Electrical-Electronics Engineering, TOBB ETÜ, Ankara, Turkiye}
\\
		{Email: zhenyuli@kth.se, ozlemtugfedemir@etu.edu.tr, emilbjo@kth.se, cavdar@kth.se }
}

\thanks{This study is conducted under the Eureka Celtic Project RAI-6Green: Robust and AI Native 6G Green Mobile Networks and partly supported by Swedish Wireless Innovations Center: SweWIN (2023-00572), both funded by Swedish Innovation Agency Vinnova.}
}



\maketitle

\begin{abstract}
    This paper investigates the application of reconfigurable intelligent surfaces (RISs) to improve fronthaul link survivability in cell-free massive MIMO (CF mMIMO) systems. To enhance the fronthaul survivability, two complementary mechanisms are considered. Firstly, RIS is set to provide reliable line-of-sight (LOS) connectivity and enhance the mmWave backup link. Secondly, a resource-sharing scheme that leverages redundant cable capacity through neighboring master access points (APs) to guarantee availability is considered. We formulate the redundant capacity minimization problem as a RIS-assisted multi-user MIMO rate control optimization problem, developing a novel solution that combines a modified weighted minimum mean square error (WMMSE) algorithm for precoding design with Riemannian gradient descent for RIS phase shift optimization. Our numerical evaluations show that RIS reduces the required redundant capacity by $65.6\%$ compared to the no RIS case to reach a $99\%$ survivability. The results show that the most substantial gains of RIS occur during complete outages of the direct disconnected master AP-CPU channel. These results demonstrate RIS's potential to significantly enhance fronthaul reliability while minimizing infrastructure costs in next-generation wireless networks.
    
\end{abstract}


\section{Introduction}

    The increasing capacity demands of mobile user equipments (UEs) necessitate network densification and bandwidth expansion to meet future requirements~\cite{jiang2021road}. In this context, cell-free massive multiple-input multiple-output (CF mMIMO) systems have emerged as a promising solution for next-generation mobile networks~\cite{kassam2023review}. However, as highlighted in~\cite{bjornson2020scalable}, the scalability of CF mMIMO systems is hindered by their high fronthaul capacity requirements. To address this, wired connections, which offer greater capacity and stability than wireless links, are primarily employed to interconnect system entities.

    Nevertheless, CF mMIMO relies on a densely distributed network of access points (APs) connected to a central processing unit (CPU) via fronthaul links, leading to potentially prohibitive cabling costs in a fully connected topology. Recent efforts have focused on optimizing network topology to reduce deployment expenses~\cite{7996771,alabbasi2018optimal,ericsson2019radiostripes}. For instance, Ericsson’s radio stripes concept~\cite{ericsson2019radiostripes} proposes a serialized cable connection for APs, minimizing the wiring and improving cost efficiency. However, such streamlined connectivity also increases vulnerability to link failures. Thus, ensuring network survivability in the event of fronthaul disruptions is critical. 

    Without deploying additional infrastructure, wireless fronthaul links can serve as a cost-effective backup solution by leveraging the existing signal transceiver capabilities of APs in the event of primary wired connection failures. However, as noted in~\cite{townend2023challenges}, wireless fronthaul faces significant challenges, including limited channel capacity compared to wired alternatives and higher sensitivity to environmental fluctuations. These limitations are more critical in urban settings, where guaranteeing consistent line-of-sight (LOS) conditions is often impractical. Reconfigurable intelligent surfaces (RISs) have emerged as a promising solution to mitigate these challenges, leveraging their ability to manipulate the wireless propagation environment while maintaining low deployment costs~\cite{zhang2021reconfigurable}. While RIS-assisted CF mMIMO systems have been extensively studied for improving access link performance, existing research has primarily focused on quality-of-service enhancements. For example,~\cite{xie2020multiple} and~\cite{huang2021towards} present joint optimization frameworks for RIS phase-shifts and AP beamformers under perfect and imperfect channel state information (CSI), respectively.

    However, the potential of RIS to enhance network survivability remains largely unexplored. Unlike conventional network components, RISs impose no fronthaul requirements and can maintain reliable operation even during link failures. Furthermore, since AP and CPU locations are typically fixed, RIS control mechanisms can be significantly simplified. Motivated by these advantages, this work investigates RIS-assisted survivable fronthaul design for CF mMIMO systems. The main contributions of this work is summarized as:
    \begin{itemize}
        \item We consider a novel fronthaul design to survive the primary cable link failure. In the designed system, RIS is set to work jointly with the disconnected master AP to allocate fronthaul load to the CPU radio head and its nearest master AP using the mmWave signal.
        \item To minimize the redundant capacity required to mitigate primary cable link failures, we propose a RIS-assisted multi-user MIMO rate control algorithm. RIS phase shifts are optimized via Riemannian gradient descent, while precoding matrices are selected through a modified weighted minimum mean square error (WMMSE)-based procedure.
        \item The numerical results have shown a significant reduction in requested redundant capacity while satisfying the same survivability level. Such reduction is revealed to be the most obvious when the RIS is assisting the channel that is in outage condition.
    \end{itemize}

\section{System model}

    \begin{figure}[tb]
        \centering
        \includegraphics[trim= 225 70 150 75, clip, width=0.9\linewidth]{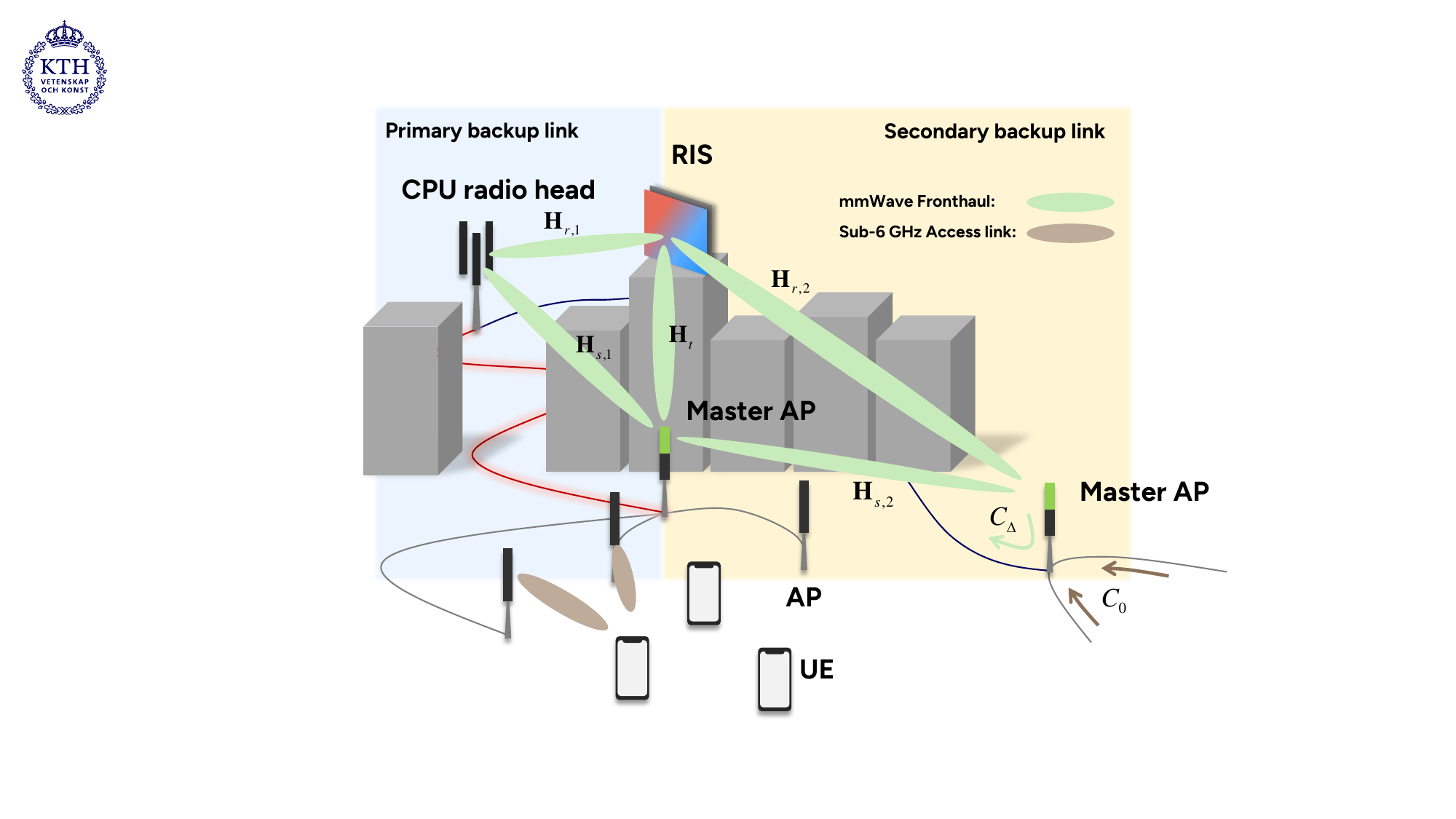}
                \vspace{-2mm}
        \caption{Illustration of the RIS-assisted CF mMIMO system model.}
        \vspace{-4mm}
        \label{fig:systemmodel}
    \end{figure}
    
    As illustrated in Fig.~\ref{fig:systemmodel}, we consider a CF mMIMO system with multiple clusters, each comprising APs and UEs, exchanging information with a shared CPU. To lower the total cabling cost, the CF mMIMO system is considered to be constructed in a tree topology~\cite{demirhan2022enabling}, where each cluster contains one master AP that aggregates data from its connected ordinary APs and transmits it to the CPU via cabled fronthaul. A wireless fronthaul link serves as a backup when the primary cable connection fails. To prevent interference between fronthaul and access links, we employ a dual-band system where access links operate in sub-6 GHz bands while fronthaul links utilize mmWave frequencies for their wider bandwidth. For conciseness, master APs experiencing cable failures are hereafter termed disconnected master APs.

    Due to human activities or environmental reasons, the primary cable connection can encounter failure. The disconnected master AP predominantly relies on its wireless link to the CPU radio head to transfer data. However, in urban environments characterized by numerous obstacles, together with the mmWave signal’s sensitivity to blockages, the predominant wireless link may fail to deliver sufficient capacity to meet the fronthaul data rate demands, particularly as propagation conditions fluctuate. To characterize this fluctuation, three distinct propagation states are modeled: outage, LOS, and non-line-of-sight (NLOS), as described in~\cite{akdeniz2014millimeter}. In the outage state, the signal incurs infinite propagation loss, rendering direct connectivity unfeasible. In contrast, the LOS and NLOS states result in different levels of attenuation. The large-scale attenuation is defined as
    \begin{equation}
        \rho = -a - 10b\log_{10}(d/d_0)
    \end{equation}
    where $d$ and $d_0$ are the propagation and the reference distance. Coefficients $a$ and $b$ take different values for LOS and NLOS conditions. The small-scale attenuation $\boldsymbol{\beta}$ also takes different distributions according to the propagation condition. For the NLOS condition, independent and identically distributed (i.i.d.) Rayleigh fading is considered, while Rician fading with the Rician factor $\kappa$ is considered for the LOS condition. Given that, an arbitrary channel can be expressed as $\mathbf{H}=\sqrt{\rho}\boldsymbol{\beta}$. The probabilities of the channel to fall into outage, LOS, and NLOS, namely $\mathcal{P}_\text{out}$, $\mathcal{P}_\text{LOS}$, and $\mathcal{P}_\text{NLOS}$, are given as \cite{akdeniz2014millimeter}
    \begin{align}
        &\mathcal{P}_\text{out}(d)=\max(0,1-e^{-a_\text{out}d+b_\text{out}}),
        \label{eq:out}\\
        &\mathcal{P}_\text{LOS}(d) = (1-\mathcal{P}_\text{out}(d))e^{-a_\text{LOS}d},\label{eq:los}\\
        &\mathcal{P}_\text{NLOS}(d) = 1-\mathcal{P}_\text{LOS}(d)-\mathcal{P}_\text{out}(d),\label{eq:nlos}
    \end{align}
    where $a_\text{out}$, $b_\text{out}$, and $a_\text{LOS}$ are frequency-related coefficients.
    
    An increase in the distance between the CPU radio head and the disconnected master AP elevates the probability of falling into a worse condition and extending the propagation attenuation of the signal, potentially resulting in insufficient channel capacity. To mitigate such conditions, we propose two complementary mechanisms. First, we deploy an RIS positioned to maintain mutual LOS with both the transmitters and receivers within its coverage area, thereby overcoming worse propagation conditions. Second, leveraging the short inter-AP distance, a resource-sharing mechanism is applied, enabling the disconnected master AP to transfer its fronthaul traffic to the nearest master AP with a wireless connection. As the reliability of the cable connection is generally high, we assume multiple cable failures are not likely to happen at the same time. Given that, this neighboring master AP subsequently processes the received signal and forwards it to the target CPU via its functional cabled connection. To ensure this additional transmission load does not disrupt normal operations, the cable must incorporate redundant capacity. We define the designed cable capacity between any master AP and the CPU as $C_\text{cable} = C_0 + C_\Delta$, where $C_0$ is the fronthaul capacity required for ordinary operating and $C_\Delta$ denotes the redundant capacity.
    
    We refer to the cascaded disconnected master AP-RIS-CPU radio head channel as the \textit{primary backup link}, and the disconnected master AP-RIS-nearest master AP channel as the \textit{secondary backup link}. For notational convenience, we use the index $k\in\{1,2\}$ to distinguish between two potential receivers: the CPU radio head when $k = 1$, and the nearest master AP when $k = 2$. The radio head of the CPU is mounted with $N_1 = N_\text{CPU}$ antennas, and the master APs are equipped with $N_2 = N_\text{AP}$ antennas for the mmWave fronthaul link. Additionally, the RIS is equipped with $M$ reflecting elements. The channel between the receiver $k$ and the disconnected master AP is denoted as $\mathbf{H}_{s,k}\in\mathbb{C}^{N_k\times N_\text{AP}}$. Moreover, $\mathbf{H}_{r,k}\in\mathbb{C}^{N_k\times M}$ and $\mathbf{H}_t\in\mathbb{C}^{M\times N_\text{AP}}$ are the channels from RIS to receiver $k$ and the disconnected master AP to RIS. For simplicity, in this work, we assume the CSI is perfectly known by all entities.

    \subsection{RIS-assisted wireless fronthaul channel}
        
        The disconnected master AP is simultaneously offloading data to the CPU and its nearest master AP. The received signal $\mathbf{y}_k\in\mathbb{C}^{N_k}$ at the receiver $k$ is given as
        \begin{equation}
            \mathbf{y}_k = (\underbrace{\mathbf{H}_{s,k}+\mathbf{H}_{r,k}\boldsymbol{\Phi}\mathbf{H}_{t}}_{\triangleq \mathbf{H}_k\in\mathbb{C}^{N_k\times N_\text{AP}}})\sum_{i=1}^2\mathbf{W}_i\mathbf{s}_i + \mathbf{n}_k        
        \end{equation}
        where $\mathbf{n}_k\sim\mathcal{CN}(\mathbf{0},BN_0\mathbf{I}_{N_k})$ is the additive white complex Gaussian noise (AWGN) vector with power spectral density $N_0$. $B$ is the available bandwidth and $\mathbf{s}_i\in\mathbb{C}^{N_\text{AP}}$ is the transmit data symbol for which $\mathbb{E}\{\mathbf{s}_i\mathbf{s}_i^H\}=\mathbf{I}_{N_\text{AP}},~\forall i$ holds. The RIS phase-shift matrix $\boldsymbol{\Phi}=\text{diag}(\boldsymbol{\phi})\in\mathbb{C}^{M\times M}$, where the RIS phase-shift vector is given as $\boldsymbol{\phi}=\begin{bmatrix}
            e^{j\varphi_1} & \cdots & e^{j\varphi_M}
        \end{bmatrix}^T\in\mathbb{C}^{M}$ with $\varphi_m\in[0,2\pi)~\forall m$. The precoding matrix is denoted as $\mathbf{W}_k\in\mathbb{C}^{N_\text{AP}\times N_\text{AP}}$, and the transmit power is constrained as 
        \begin{equation}
            \sum_{k=1}^2 \text{Tr}\left(\mathbf{W}_k\mathbf{W}_k^H\right)\leq P.\label{eq:powercon}
        \end{equation}
        The corresponding individual rate for receiver $k$ is given as
        \begin{align}
            R_k = &B\log_2\det\Bigg(\mathbf{I}_{N_k}+\mathbf{H}_k\mathbf{W}_k\mathbf{W}_k^H\mathbf{H}_k^H\cdot\notag\\
            &\Bigg(\sum_{i\neq k}\mathbf{H}_k\mathbf{W}_i\mathbf{W}_i^H\mathbf{H}_k^H+BN_0\mathbf{I}_{N_k}\Bigg)^{-1}\Bigg)\label{eq:individualrate}.
        \end{align}

    \subsection{Survivability of the fronthaul connection}

        Variations in channel conditions and fluctuations in the small-scale fading necessitate adaptive transmission strategies, such as reconfiguring RIS phase shifts and adjusting the precoding matrix to redistribute loads. For instance, a transition from NLOS to outage in the channel between a disconnected master AP and the CPU radio head drastically reduces capacity, requiring load reallocation to the nearest master AP to maintain the target sum rate. Following~\cite{johnston2014robust}, the fronthaul survivability $\epsilon\in[0,1]$ is defined as
        \begin{equation}
            \epsilon = \mathbb{P}\{R_1+R_2 \geq C_0| C_\Delta\}\label{eq:survivability}
        \end{equation}
        which is the probability of the total backup capacity suffices under primary operational demands, given a predetermined redundant capacity design.
    
\section{Redundant capacity optimization}

    To satisfy a higher capacity requirement, a cable with a higher capacity is needed. Consequently, the deployment cost will also increase. This cost escalation becomes particularly significant in CF mMIMO systems due to the dense distribution of the infrastructure. Therefore, optimizing redundant capacity requested is crucial for minimizing overall deployment cost. For that purpose, a redundant capacity minimization problem $\textbf{P1}$ is formulated as
    \begin{subequations}
        \begin{align}
            \textbf{P1}: \quad&\underset{\{\mathbf{W}_k\},  \boldsymbol{\Phi}}{\text{minimize}} \quad C_\Delta \label{eq:objective}\\
            &\text{s.t.} \quad R_1+R_2 \geq C_0, \label{eq:ratecon}\\
            &\hspace{7.5mm} R_2 \leq C_\Delta, \label{eq:capacity}\\ 
            &\hspace{7.5mm} \eqref{eq:powercon}. \notag
        \end{align}
    \end{subequations}

    The objective \eqref{eq:objective} is to minimize the redundant capacity designed for the primary cable connection by jointly selecting the RIS phase shifts and the precoding matrices. Constraint \eqref{eq:ratecon} ensures that the sum rate satisfies the fronthaul capacity requirement and~\eqref{eq:capacity} ensures the requested rate of the secondary backup link does not exceed the given redundancy. Moreover, \eqref{eq:powercon} ensures the transmit power does not exceed $P$. 

    Notice that, due to the quadratic variable coupling and the non-convex fractional structure of the RIS phase shifts and precoding matrices in \eqref{eq:individualrate}, the objective \eqref{eq:objective} and constraint \eqref{eq:ratecon} violate the convexity, which renders $\textbf{P1}$ complicated to be solved to global optimality. To tackle this, an alternating optimization approach is proposed to solve $\textbf{P1}$. Firstly, as the maximum of $R_2$ is limited by $C_\Delta$, optimality will be achieved when $C_\Delta=R_2$, thus, the objective can be equivalently replaced by $R_2$, and removing~\eqref{eq:capacity}. Additionally, we introduce Lagrange multiplier $\lambda$, and denote the Lagrangian function with~\eqref{eq:powercon} neglected as 
    \begin{equation}
        \mathcal{L}(\{\mathbf{W}_k\},\boldsymbol{\Phi},\lambda)=R_2+\lambda(C_0-R_1-R_2).
    \end{equation}
    The Lagrangian dual problem of $\textbf{P1}$ is
        \begin{equation}
            \textbf{P2}: \quad \underset{\lambda}{\text{maximize}}~\underset{\{\mathbf{W}_k\},  \boldsymbol{\Phi}}{\text{min}} \quad \mathcal{L}(\{\mathbf{W}_k\},\boldsymbol{\Phi},\lambda).
        \end{equation}
    The neglected power constraint \eqref{eq:powercon} is guaranteed later with the modified WMMSE-based procedure.
        
    \subsection{Modified WMMSE-based precoding optimization}
    In this part, we optimize the precoding matrices for fixed RIS phase-shift configuration and $\lambda$. If $\lambda\leq 1$, the problem becomes maximizing $R_1$ while setting $\mathbf{W}_2=\mathbf{0}$. This problem can be solved with the classical water-filling algorithm. When $\lambda>1$, we obtain a weighted sum rate problem. The WMMSE-based precoding selection approach is widely used in multi-user MIMO systems to maximize the sum rate~\cite{christensen2008weighted}. The precoding matrices are determined to minimize the mean square error (MSE) of the received signal. The MSE matrix for receiver $k$ is denoted as
        \begin{equation}
            \mathbf{E}_k = \mathbb{E}\left\{(\mathbf{s}_k - \mathbf{U}_k\mathbf{y}_k)(\mathbf{s}_k - \mathbf{U}_k\mathbf{y}_k)^H\right\}.\label{eq:MSEorg}
        \end{equation}
        where $\mathbf{U}_k\in\mathbb{C}^{N_\text{AP}\times N_k}$ is the MSE equalizer. The optimal $\mathbf{U}_k$ that minimizes the trace of  the MSE matrix is given as
        \begin{equation}
            \mathbf{U}_k = \mathbf{W}_k^H\mathbf{H}_k^H(\mathbf{H}_k\mathbf{WW}^H\mathbf{H}_k+BN_0\mathbf{I}_{N_k})^{-1}
        \end{equation}
        where $\mathbf{W}=\begin{bmatrix} \mathbf{W}_1 &\mathbf{W}_2 \end{bmatrix}\in\mathbb{C}^{N_\text{AP}\times (N_\text{AP}+N_\text{CPU})}$ is the concatenated precoding matrix. Substituting it back into \eqref{eq:MSEorg}, the MSE matrix can be reformatted as
        \begin{equation}
            \mathbf{E}_k=(\mathbf{I}_{N_k}+\mathbf{W}_k^H\mathbf{H}_k^H\mathbf{Q}_k^{-1}\mathbf{H}_k\mathbf{W}_k)^{-1},
        \end{equation}
        where $\mathbf{Q}_k=\mathbf{H}_k\sum_{i\neq k}\mathbf{W}_i\mathbf{W}_i^H\mathbf{H}_k^H+BN_0\mathbf{I}_{N_k}$. Instead of directly choosing the MSE weight matrices $\mathbf{V}_k\in\mathbb{C}^{N_\text{AP}\times N_\text{AP}}$ to maximize the sum rate as in~\cite{christensen2008weighted}, we scale the MSE weight matrices as
        \begin{align}
            &\mathbf{V}_1 = \lambda \mathbf{E}_1^{-1},\quad \mathbf{V}_2 = (\lambda-1) \mathbf{E}_2^{-1},
        \end{align}
        and the corresponding precoding matrices are given as
        \begin{equation}
            \mathbf{W}_k = \left(\sum_{i=1}^2\mathbf{H}_i^H\mathbf{U}_i^H\mathbf{V}_i\mathbf{U}_i\mathbf{H}_i+\mu\mathbf{I}_{N_\text{AP}}\right)^{-1}\mathbf{H}_k^H\mathbf{U}_k^H\mathbf{V}_k\label{eq:precoding}
        \end{equation}
        where the auxiliary non-negative variable $\mu$ is chosen to meet the power constraint. With perfect CSI and given RIS phase shifts, constraint~\eqref{eq:powercon} exhibits a monotonic relationship with $\mu$, enabling efficient computation of $\mu$ via bisection search.  
        

    \subsection{RIS phase shifts optimization}

        The coupling between the desired signal and interference terms presents a significant challenge in isolating the term that is coupled to $\phi_m$. This prevents the direct application of conventional block coordinate descent (BCD) methods~\cite{zhang2020capacity}. Considering the phase shifts are on the unit circle, Riemannian gradient descent is utilized. Without deviating the objective in $\textbf{P2}$, the RIS is designed to work in a way that minimizes $\mathcal{L}$ by adjusting $\boldsymbol{\phi}$ over the manifold $\mathcal{M}=\{\boldsymbol{\phi}\in\mathbb{C}^M|\, |\phi_m|=1,~ \forall m\}$.

        With the precoding matrices determined with the modified WMMSE-based procedure, $\mathbf{W}_k$ is considered given and independent of $\boldsymbol{\Phi}$. By applying the chain rule, the Euclidean gradient with regard to $\mathcal{L}$ over $\phi_m$ can be expressed as
        \begin{align}
            \frac{\partial \mathcal{L}}{\partial \phi_m} = (1-\lambda)\frac{\partial R_2}{\partial \phi_m} -\lambda \frac{\partial R_1}{\partial \phi_m} \label{eq:dldphi}
        \end{align}
        where $R_k$ is real and $\mathbf{H}_k$ and $\phi_m$ are complex, based on the Wirtinger derivatives as well as the chain rule
        \begin{equation}
            \frac{\partial R_k}{\partial \phi_m} = \text{Tr}\left(\left(\frac{\partial R_k}{\partial \mathbf{H}_k}\right)^H\frac{\partial \mathbf{H}_k}{\partial \phi_m^*}\right)+\text{Tr}\left(\left(\frac{\partial R_k}{\partial \mathbf{H}_k^H}\right)^H\frac{\partial \mathbf{H}_k^H}{\partial \phi_m^*}\right).\label{eq:grad}
        \end{equation}
        Notice that since $\mathbf{H}_k$ is holomorphic in $\phi_m$, thus $\frac{\partial \mathbf{H}_k}{\partial \phi_m^*} = \mathbf{0}$. As a result, $\frac{\partial R_k}{\partial \phi_m}$ is determined only by the second term. Denote $\mathbf{h}_{r,k,m}$ and $\overrightarrow{\mathbf{h}}_{t,m}$ as the $m$-th column of the $\mathbf{H}_{r,k}$ and the $m$-th row of $\mathbf{H}_{t}$ respectively. Then $\mathbf{H}_k^H$ can be reformatted as $\mathbf{H}_k^H = \mathbf{H}_{s,k}^H+\sum_{m=1}^M\phi_m^*\overrightarrow{\mathbf{h}}_{t,m}^H\mathbf{h}_{r,k,m}^H$ and the partial derivative $\frac{\partial \mathbf{H}_k^H}{\partial \phi_m^*}$ can be given as
        \begin{equation}
            \frac{\partial \mathbf{H}_k^H}{\partial \phi_m^*} = \overrightarrow{\mathbf{h}}_{t,m}^H\mathbf{h}_{r,k,m}^H. \label{eq:dh1dphi}
        \end{equation}
        To derive $\partial R_k/\partial \mathbf{H}_k$, we first reformat the $R_k$ as
        \begin{align}
            R_k = &B\log_2\det\Bigg(\mathbf{H}_k\bigg(\sum_{i=1}^2\mathbf{W}_i\mathbf{W}_i^H\bigg)\mathbf{H}_k^H + BN_0\mathbf{I}_{N_k}\Bigg)  \notag\\
            &\hspace{-2mm}-B\log_2\det\bigg(\mathbf{H}_k\sum_{i\neq k}\mathbf{W}_i\mathbf{W}_i^H\mathbf{H}_k^H + BN_0\mathbf{I}_{N_k}\bigg)
        \end{align}
        and $\partial R_k/\partial \mathbf{H}_k$ is calculated as
        \begin{align}
            \frac{\partial R_k}{\partial \mathbf{H}_k} = &\frac{B}{\ln2}\Bigg(\bigg(\sum_{i=1}^2\mathbf{W}_i\mathbf{W}_i^H\bigg)\mathbf{H}_k^H\Bigg(\mathbf{H}_k\bigg(\sum_{i=1}^2\mathbf{W}_i\mathbf{W}_i^H\bigg)\mathbf{H}_k^H\notag\\
            &+BN_0\mathbf{I}_{N_k}\Bigg)^{-1}-\sum_{i\neq k}\mathbf{W}_i\mathbf{W}_i^H\mathbf{H}_k^H\times\notag\\
            &\Bigg(\mathbf{H}_k\bigg(\sum_{i\neq k}\mathbf{W}_i\mathbf{W}_i^H\bigg)\mathbf{H}_k^H+BN_0\mathbf{I}_{N_k}\Bigg)^{-1}\Bigg)\label{eq:dr1dh1}.
        \end{align}
        Furthermore, since $R_k$ is real, $\left(\frac{\partial R_k}{\partial \mathbf{H}_k^H}\right)^H = \frac{\partial R_k}{\partial \mathbf{H}_k}$ holds. By substituting~\eqref{eq:grad}, ~\eqref{eq:dh1dphi}, and \eqref{eq:dr1dh1} to \eqref{eq:dldphi}, the Euclidean gradient can be obtained analytically. Then, the Euclidean gradient is projected to the manifold, and the Riemannian gradient is calculated as
        \begin{equation}
            [\text{grad}_\mathcal{M}\mathcal{L}]_m = \frac{\partial \mathcal{L}}{\partial \phi_m} - \Re\left(\left(\frac{\partial \mathcal{L}}{\partial \phi_m}\right)^*\phi_m\right)\phi_m.
        \end{equation}
        In terms of minimizing $\mathcal{L}$, the phase shifts are updated with Riemannian gradient descent
        \begin{equation}
            \boldsymbol{\phi} = \exp\left(j\cdot\angle(\boldsymbol{\phi} - \eta\cdot\text{grad}_\mathcal{M}\mathcal{L})\right)
        \end{equation}
        where $\eta$ is the step size, and $\angle(\cdot)$ obtains the angle of a complex value.

    \subsection{Lagrange multiplier updates}

        With $\mathbf{W}_k$ and $\boldsymbol{\Phi}$ computed, the individual rates $R_1$ and $R_2$ can be calculated correspondingly. Based on the constraint violation condition, the Lagrangian multiplier $\lambda$ is updated accordingly. The partial derivative of the Lagrangian function over the Lagrangian multiplier is $\frac{\partial \mathcal{L}}{\partial \lambda} = C_0 - R_1-R_2$ then $\lambda$ is updated in a gradient ascent manner with a step size $\alpha$ as
        \begin{equation}
            \lambda = \lambda + \alpha (C_0-R_1-R_2).
        \end{equation}
        When the current sum rate exceeds the required data rate, $\lambda$ decreases, resulting in the modified WMMSE procedure attempting to reduce $R_2$. On the contrary, when the current sum rate is lower than the required data rate, $\lambda$ increases, resulting in more power allocated to the secondary backup link, where the channel capacity is typically higher.

    \subsection{RIS-assisted multi-user MIMO system rate control algorithm}

        Algorithm~\ref{alg:ratecontrol} outlines the RIS-assisted multi-user MIMO rate control procedure. By introducing the Lagrangian multiplier $\lambda$, the constrained problem $\textbf{P2}$ becomes unconstrained, avoiding infeasibility. For each $\lambda$, the precoding matrices $\mathbf{W}_k$ are computed in closed form with~\eqref{eq:precoding}, and the RIS phase shifts are optimized iteratively. The algorithm converges as the bounded $\lambda$ reaches a value where the sum rate meets $C_0$ or its boundary.

      \section{Numerical results}

\begin{figure}[tb!]
        \centering
        \includegraphics[trim= 160 150 170 150, clip, width=0.9\linewidth]{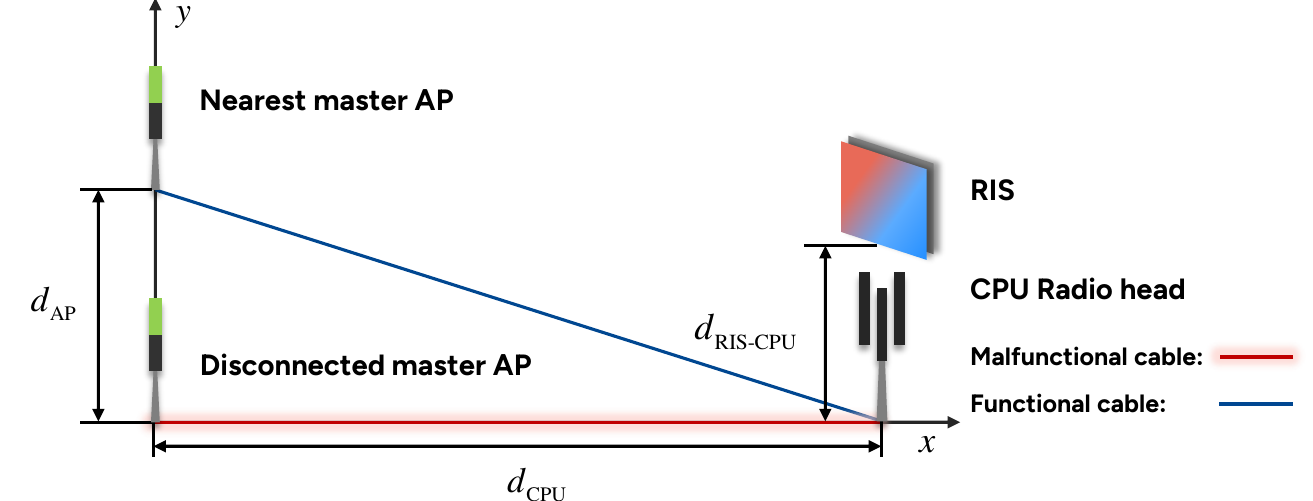}
                        \vspace{-3mm}
        \caption{Illustration of the simulation setup. The height of different entities is neglected for simplicity.}
        \label{fig:simulationsetup}

    \end{figure}

    We consider the simulation setup as illustrated in Fig.~\ref{fig:simulationsetup}. The disconnected master AP is located at the Cartesian coordinate $(0,0)$. The position of the nearest master AP, CPU radio head, and RIS are set to be $(0, d_\text{AP})$, $(d_\text{CPU},0)$, and $(d_\text{CPU},d_\text{RIS-CPU})$. As indicated in~\cite{bjornson2024introduction}, with the purpose of better enhancing point-to-point communication, RIS should be positioned either close to the transmitter or the receiver. In our case, placing the RIS close to the disconnected master AP could enlarge the impact of the interference, which potentially lowers the RIS's performance. Given that, we consider the RIS to be placed close to the CPU. When generating channel matrices, each realization takes different channel conditions according to the probability mass function (PMF) described with~\eqref{eq:out} to~\eqref{eq:nlos}. The detailed system and channel parameters are summarized in Table~\ref{tab:parameters}.
        \begin{algorithm}
            \caption{RIS-assisted multi-user MIMO system rate control} \label{alg:ratecontrol}
            \begin{algorithmic}[1]
                \State {\bf Given:} The transmit power budget $P$. The channel matrices $\mathbf{H}_{s,k}$, $\mathbf{H}_{r,k}$ for all $k$ and $\mathbf{H}_{t}$. The data rate requirement $C_0$. The maximum rounds of iteration $T$ and $E$. The step size $\alpha$ and $\eta$.
                
                \State {\bf Initialization:}  Initialize the concatenated precoding matrix $\mathbf{W}^{(0)} = \sqrt{P}\mathbf{H}/\|\mathbf{H}\|_F$, where $\mathbf{H} = \begin{bmatrix}
                    \mathbf{H}_1 & \mathbf{H}_2
                \end{bmatrix}$. Initialize $\phi_{m}^{(0)}$ randomly while keeping $|\phi_{m}^{(0,E)}|=1, ~ \forall m$. Set the Lagrangian multiplier $\lambda^{(0)} = 1$
                \For {$t = 1 : T$}
                    \State $\mathbf{W}^{(t)}$ = \textit{\textbf{modifiedWMMSEPrecoding}}($\lambda^{(t-1)}$,$\boldsymbol{\Phi}^{(t-1,E)}$, $\mathbf{H}_{k}$, $P$, $C_0$)  
                    \For {$e = 1 : E$}
                        \State \hspace{-3mm}$\boldsymbol{\Phi}^{(t,e)}$ = \textit{\textbf{riemannianGradientDescent}}($\mathbf{W}^{(t)}$, $\mathbf{H}_{k}$, $\eta$)
                    \EndFor 
                    \State $\lambda^{(t)}$ = \textit{\textbf{lagrangianMultiplierUpdating}}($\mathbf{W}^{(t)}$,$\boldsymbol{\Phi}^{(t,E)}$,$\alpha$,  $C_0$)
                    \State $\lambda^{(t)}$ = $\max(1, \lambda^{(t)})$
                \EndFor
                \State{\textbf{Output:}} The RIS phase-shift matrix $\boldsymbol{\Phi}^{(T,E)}$. Precoding matrices $\mathbf{W}_k^{(T)}$. Lagrange multiplier $\lambda^{(T)}$.
            \end{algorithmic}
        \end{algorithm}

    \begin{table}[t]
    \caption{Key System and Channel Parameters}
    \label{tab:parameters}
    \centering
    \begin{tabular}{@{}lll@{}}
        \toprule
        \textbf{Category} & \textbf{Parameter} & \textbf{Value} \\
        \midrule
        Carrier frequency & \( f_c \) & 28 GHz \\
        Bandwidth & \( B \) & 200 MHz \\
        Transmit power budget & \( P \) & 10 W \\
        Master AP antennas & \( N_{\text{AP}} \) & 32 \\
        CPU radio head antennas & \( N_{\text{AP}} \) & 32 \\
        RIS elements & \( M \) & 1024 \\
        Rician factor & \( \kappa \) & 10 \\
        Pathloss coefficients & \multicolumn{2}{l}{See Table I in~\cite{akdeniz2014millimeter}} \\
        \bottomrule
    \end{tabular}
                    \vspace{-2mm}
\end{table}

    The fronthaul data rate requirement is determined via functional splitting scheme 7.2 as~\cite{perez2018fronthaul}
    \begin{equation}
        C_0=\frac{2N_\text{used}N_\text{bit}N_\text{ac}}{T_s},
    \end{equation}
    where $N_\text{used}$ is the number of used subcarriers and $T_s$ is the OFDM symbol duration. Additionally, $N_\text{ac}$ is the number of antennas used for access link transmission, and $N_\text{bit}$ is the quantization bit per symbol. In this work, we consider $N_\text{ac} = 12$, $T_s = 71.4$ $\mu$s, and $N_\text{bits}=12$ bit per symbol \cite{demir2024cell}. 

\begin{figure}[tb!]
        \centering
        \includegraphics[width=0.8\linewidth]{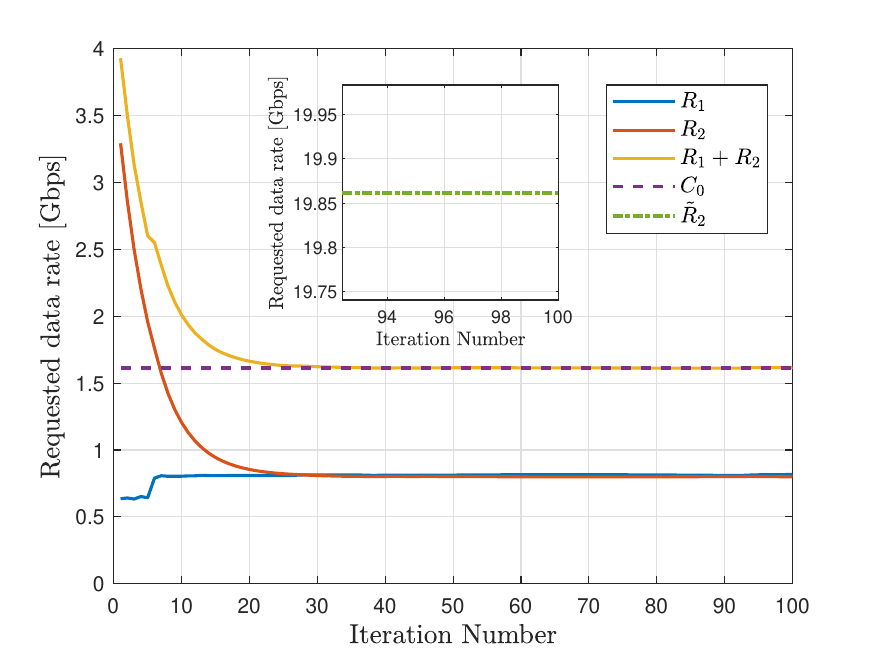}
                        \vspace{-3mm}
        \caption{Convergence condition of the algorithm. In the evaluated case, $N_\text{used}=400$, $C_0=1.6$ Gbps, $d_\text{AP}=50$ m, $d_\text{CPU}=200$ m, and $d_\text{RIS-CPU}=5$ m.}
        \label{fig:convergence}
                        \vspace{-4mm}
    \end{figure}

    To evaluate the convergence of our proposed algorithm, the algorithm is run by inputting one realization of the channel. There, the maximum iteration $E=50$ for the RIS phase-shifts updating and $T=100$ for the algorithm. As depicted in Fig.~\ref{fig:convergence}, the sum rate converged to $C_0$ within approximately 30 rounds. Additionally, to show that our proposed algorithm is minimizing $R_2$, an extra simulation is done with the conventional WMMSE and the assistance of RIS, with which the sum rate is expected to be maximized. Given that, the achieved data rate of the secondary backup link, denoted as $\tilde{R}_2$, is observed to be significantly higher than $R_2$.

\begin{figure}[tb]
        \centering
        \includegraphics[width=0.9\linewidth]{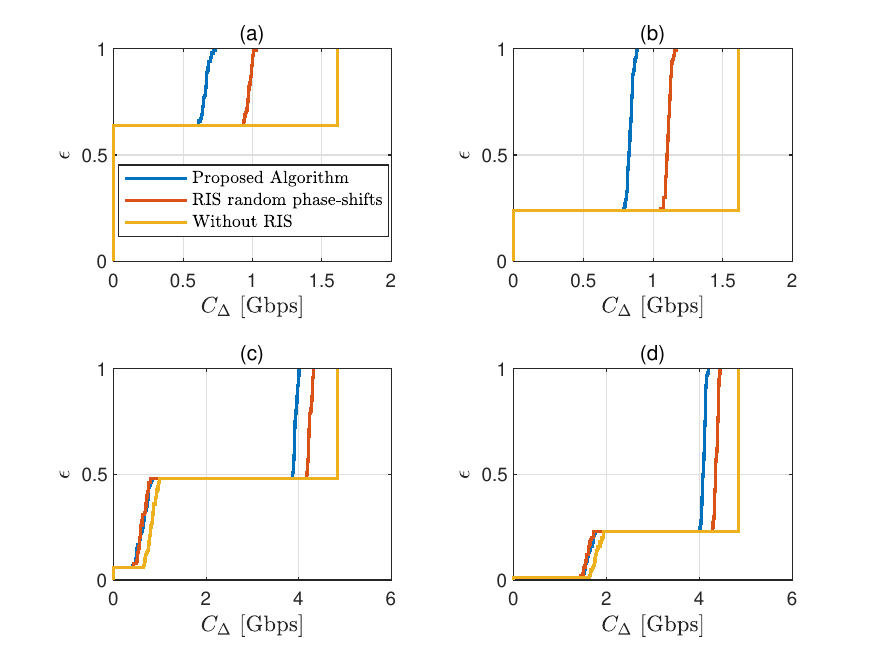}
                        \vspace{-3mm}
        \caption{Relation between the redundant capacity and the resulting survivability. (a) $d_\text{CPU}=175$ m, $N_\text{used}=400$, $C_0=1.6$ Gbps; (b) $d_\text{CPU}=200$ m, $N_\text{used}=400$, $C_0=1.6$ Gbps; (c) $d_\text{CPU}=175$ m, $N_\text{used}=1200$, $C_0=4.9$ Gbps; (d) $d_\text{CPU}=200$ m, $N_\text{used}=1200$, $C_0=4.9$ Gbps.}

        \label{fig:results}
                        \vspace{-4mm}
    \end{figure}

    Redundant capacity is optimized over 100 channel realizations under different $d_\text{CPU}$ and $N_\text{used}$ and is illustrated in Fig.~\ref{fig:results}.  The results indicate that RIS significantly reduces redundant capacity requirements to achieve target survivability, even without phase shifts optimization. This improvement arises from the RIS’s deployment that preserves LOS propagation paths, mitigating outages. In non-RIS scenarios, when $\mathbf{H}_{s,1}$ falls into outage, only the secondary backup link can be established, resulting in a distinct step at $C_\Delta=C_0$. Moreover, the proposed algorithm reduces redundant capacity by $65.6\%$ compared to non-RIS cases to guarantee a $99\%$ survivability. Notably, favorable channel conditions eliminate the need for resource sharing between master APs, yielding an intercept point. 
    
    Comparing Fig.~\ref{fig:results}(a) and (b), as the $d_\text{CPU}$ gets larger, the probability of the channel falling into a worse condition rises and the propagation attenuation increases, causing the intercept to decrease. Conversely, when master APs are positioned sufficiently close to the CPU, the wireless backup fronthaul achieves high survivability without needing redundant capacity. Furthermore, with the same $C_0$, the reduction brought by the RIS also decays with the extension of $d_\text{CPU}$. This is due to the higher propagation attenuation applied to the signal received by the RIS.

    Fig.~\ref{fig:results}(a) and (c) demonstrate that the performance improvement provided by the RIS diminishes to $17.2\%$ as the required fronthaul rate increases. This is because the system preferentially allocates more transmit power to enhance the secondary backup link, which offers superior channel capacity, consequently satisfying the higher rate demand. Additionally, a second step suggesting $\mathbf{H}_{s,1}$ is in NLOS condition can be observed in Fig.~\ref{fig:results}(c) and (d) as the primary backup link can no longer provide sufficient capacity for the higher rate demand. In that case, as power is more aligned to $\mathbf{H}_{s,1}$ and the secondary backup link, RIS captures less power, and its contribution to reducing the redundant capacity as a result becomes less pronounced.  
    
\section{Conclusion}

    In this paper, we explore the use of RIS to enhance the survivability of fronthaul links for CF mMIMO systems. During fronthaul cable failures, a disconnected master AP establishes a wireless mmWave backup link to the CPU. We deploy and optimize an RIS to enable LOS connectivity, improving channel reliability. A resource-sharing mechanism allows the disconnected AP to offload traffic to the nearest master AP, utilizing redundant capacity reserved for cabled links. To minimize redundant capacity, we propose an RIS-assisted MU-MIMO rate control algorithm. This algorithm reduces the rate requested with the secondary backup link while maintaining a predefined sum rate. Simulations demonstrate that RIS reduces redundant capacity requirements by up to $65.6\%$ compared to non-RIS cases to achieve a $99\%$ survivability. Furthermore, our findings reveal that the performance gains offered by RIS are most pronounced when the direct channel between the disconnected master AP and the CPU experiences an outage.
        
\bibliographystyle{IEEEtran}

\bibliography{Main}

\end{document}